# Positron lifetime calculations for defects in Zn

J.M. Campillo [1], F. Plazaola [2] and N. de Diego [3]

(1) Oinarrizko Zientziak Saila, Goi Eskola Politeknikoa, Mondragon Unibertsitatea, 20500 Arrasate, Spain
(2) Elektrika eta Elektronika Saila, Zientzi Fakultatea, UPV-EHU, 48080 Bilbo, Spain
(3) Dpto. de Física de Materiales, Facultad de Ciencias Físicas, Universidad Complutense, 28040 Madrid, Spain

Short title:     Positron lifetime calculations in defects in Zn

[1] jmcampillo@eps.muni.es
[2] fernando@we.lc.ehu.es
[3] nievesd@eucmos.sim.ucm.es

**Abstract**

The effect of the lattice relaxation at vacancy clusters and interstitial-type dislocation loops on the positron lifetime in Zn has been studied. Defective relaxed structures for the lifetime calculations have been generated by using a many-potential for Zn. From the results it is inferred that the effect of the to atomic relaxation is mainly significant for small vacancy clusters. The lifetime associated to interstitial-type loops is very sensitive to the loop structure and its surroundings. Previous experimental results are compared with the theoretical calculations.

# 1. Introduction

The theoretical methods to calculate positron lifetimes in solids are nowadays well established and several studies have been carried out covering a wide range of materials [1]. Generally the lifetime calculations do not take into account the atomic relaxations around defects, that would be expected to influence the positron lifetime, and only a few works have been devoted to study this effect on the positron lifetime. The effect of the atomic relaxation on the lifetime associated to vacancies in Mo was discussed by Hansen et al. [2]. It is also expected to be important in defects such as dislocation loops and it has been taken into account by Häkkinen, Mäkinen and Manninen [3], who calculated the lifetimes associated to edge dislocations and dislocation loops in Al and Cu in the defective crystal structures obtained by molecular dynamics. Later, Khanna et al. [4] examined the influence of relaxation on the positron lifetimes associated to vacancy clusters in Mo and they found that considering atomic relaxations around monovacancies brings the theoretical values closer to the experiment but the effect is less pronounced for large three-dimensional clusters. More recently, Kamimura et al. have reported a difference in the lifetime values for the unrelaxed and relaxed vacancy in Fe [5]. In this work the positron lifetimes were performed in relaxed defective structures obtained by using empirical many-body potentials and the same method was used to study the positron properties in dislocations in Fe [6] and in Ni [7]. Strictly, the lattice relaxation should be accomplished in the presence of the positron but the aforementioned procedure can give an idea of the extent of the effect. Thus, it seems reasonable to apply the current lifetime calculation methods to the relaxed defective crystals, in order to obtain theoretical data available to compare with experimental results. However, the main drawback of this approach is the proper description of the crystal lattice by an interatomic potential suitable for computer simulation to generate the defect structures. Recently, an empirical potential for Zn has been developed that describes satisfactorily some point defect properties [8,9] and we have used it to calculate the positron lifetimes associated to several types of defects and compare the results with the existing experimental positron data.

## 2. Calculation method

### 2.1 Relaxed models

We have simulated Zn by using the many-body potential derived by Mikhin and de Diego [8]. It is based on the second moment of the tight-binding approximation and describes satisfactorily some important physical properties such as the *c/a* ratio, cohesive energy, vacancy formation energy and elastic constants (except $C_{33}$). It demonstrates a strong anisotropy of vacancy migration with preferential jumping on the basal plane in agreement with the experiment and gives an improved value of the stacking fault energy in comparison with previous potentials. Moreover, the properties of single interstitial and interstitial clusters derived from the model explain satisfactorily some experimental observations concerning the formation of interstitial-type dislocation loops [10]. Details on the potential development and defect properties are given elsewhere [8,9].

To create the crystal containing the defect, we generated a crystallite containing 9800 relaxable atoms and introduced the necessary number of vacancies or interstitials to form the unrelaxed defect structure. The atomic positions in the relaxed configuration were obtained by carrying out calculations at constant volume and the conjugate gradients method was used to achieve the relaxed state of minimum energy in the defective crystal. The blocks containing the relaxed atom coordinates have been incorporated into the models for the lifetime calculations.

Crystal blocks containing a mono-, a di-, and a tri-vacancy in the basal plane have been generated. Small voids containing 5, 13, 26 and 57 vacancies were also constructed. The 13- and 57-vacancy clusters correspond to approximately spherical volumes of radii *a* and 2*a* respectively, *a* being the lattice parameter. It is to be remarked that volume clusters are found to have the highest binding energy with the potential model used and thus they are preferred in comparison to planar clusters. We will only consider the former in our calculations. In Fig. 1 basal plane projection has been plotted to show the vacancy configurations used to produce the defective crystals. The two non-equivalent tri-vacancy structures have been labelled as (a) and (b); the 5-vacancy cluster, not shown in the figure, is formed from the tri-vacancy labelled (a) by removing its two nearest atoms placed at $\pm c/2$.

There are several possible configurations that can give rise to interstitial-type loops in hcp crystals but we have selected only those that are the most favoured in the potential model used. The planar clusters formed by O-type interstitials (the notation for the self interstitial sites is that currently used and introduced by Johnson and Beeler [11]) result in the most stable configurations among all the possible that can be generated by considering all the favourable cases [9]; it has been found that these planar agglomerates generate faulted interstitial-type loops with $\mathbf{b}=1/2\langle 0001\rangle$, in agreement with transmission electron microscope observations [10]. Unrelaxed configurations formed by two adjacent layers of interstitials can relax to *double* loops. The most favoured configurations are formed by O-T and O-S pairs, that generate faulted double loops, and S-T and S-S pairs, that give rise to unfaulted double loops [9]. In the light of these results, we have chosen for our study the three most favoured interstitial-type loops, i.e., O-type faulted single loop, O-T-type faulted double loop and S-S-type unfaulted double loop.

As far as the single loop is concerned and in order to evaluate the possible dependence of the loop size on the lifetime value, two loops containing 7 and 19 O-type defects lying in the same plane were studied. Their scheme is shown in Fig. 2(a) in a basal plane projection. Fig. 2(b) shows a $\langle 1\bar{1}00\rangle$-type projection of the relaxed configuration of a 19-O loop. To study the effect of a vacancy located near the loop we have also generated two blocks with vacancies located in first and second neighbour positions (see positions (1) and (2) respectively in fig. 2(b)). To obtain the final configuration in these cases we have built the block containing the relaxed loop structure plus the vacancy and have allowed the whole to relax.

To study the double loops we have chosen the relaxed configurations formed by 64 S-S and by 64 O-T pairs. They generate an unfaulted and faulted double loop respectively. The $\langle 1\bar{1}00\rangle$-type projections of the two structures are shown in Figs. 3(a) and (b). It is recalled that faulted and unfaulted double loops have been observed by transmission electron microscopy in electron irradiated Zn [10].

The positron lifetime has been calculated for each of the model structures by applying the method that will be described in the next section.

## 2.2 Positron lifetime calculations

The positron lifetime calculations have been performed by using the atomic superposition method of Puska and Nieminen [12]. Even though this method is a simple one that makes use of non-self-consistent unrelaxed electronic densities, it gives satisfactory values for lifetimes in simple metals [13,14]. The good agreement between the experimental and theoretical lifetimes is mainly due to the fact that the positron annihilation rate is obtained as an integral over the product of positron and electron densities. The positron density relaxes following the electron charge transfer, keeping the value of the overlap integral constant. For this reason the lifetime calculations are not too sensitive to self-consistency. Moreover, the agreement with experimental bulk and monovacancy lifetimes has been further improved by the density gradient correction scheme [15,16]. Besides, the atomic superposition method is much less time consuming than self-consistent methods, and it is very profitable for low symmetry or extended defects such as vacancy clusters where the use of self-consistent methods is prohibitive.

The potential felt by the positron in the solid, $V_+(\mathbf{r})$, has been obtained as

$$V_+(\mathbf{r}) = V_C(\mathbf{r}) + V_{corr}(n_-(\mathbf{r}))$$

where $n_-(\mathbf{r})$ is the electron density, $V_C(\mathbf{r})$ is the Coulomb potential due to the nuclei and the electron density and $V_{corr}(\mathbf{r})$ is the positron-electron correlation energy. The positron annihilation rate has been obtained from the overlap of positron and electron density as

$$\lambda = \pi r_0^2 c \int d\mathbf{r}\, n_+(\mathbf{r})\, n_-(\mathbf{r})\, \gamma(\mathbf{r})$$

where $r_0$ is the classical electron radius, $c$ is the speed of light, $n_+(\mathbf{r})$ is the positron density and $\gamma(\mathbf{r})$ is the so-called enhancement factor. $V_{corr}(n)$ and $\gamma(\mathbf{r})$ have been taken into account by two different schemes:

1) within the LDA the interpolation formulae by Boronski and Nieminen [17] based on the results by Arponen and Pajanne [18] are used for the correlation energy, and for the enhancement the widely used form of Ref. 17, that is based on Lantto's hypernetted chain approximation calculations [19].

2) within the GGA the correlation energy and the enhancement factor due to Barbiellini et al. [15] is used. These are both based on the results by Arponen and Pajanne [18].

In the following, the results obtained within the first and second scheme will labelled BN and GGA respectively.

We have used the supercell approximation to calculate the positron potential at the nodes of a three-dimensional mesh. The Schrödinger equation is constructed in a discretised form, and the positron wave function and the positron energy eigenvalues are solved iteratively in a real space mesh [20]. For the determination of the positron wave function periodic boundary conditions have been used.

For each defect, the supercells were built with increasing size in order to insure the convergence. All the relaxed atoms were taken inside the largest supercell and their total number depended on the studied defect, 1600 being the highest. In the calculations we have used a cubic mesh with a density of 3.2 points per atomic unit in each direction.

## 3. Results and discussion

The positron lifetimes for bulk Zn and unrelaxed monovacancy have been previously calculated by using the BN and the GGA models [21]. Table 1 shows these values, together with the di- and trivacancy theoretical values and the known experimental ones. The value $\Delta\tau=\tau_u-\tau_r$ denotes the difference between the lifetime in the unrelaxed and relaxed defect. The relaxation experienced by the first and second neighbours, calculated with respect to the defect centre and obtained from the potential model used, is also shown in the table. It is clearly indicated that the theoretical positron lifetimes obtained under the BN model are quite far from the experimental values whereas the GGA approximation gives a better fit to the experiment. Therefore, we conclude that the lifetimes calculated within the GGA approximation are more suitable and we will refer only to them in the following. It is however to be pointed out that, to our knowledge, only monovacancies have been treated within the GGA approximation. There is no experience in the case of vacancy clusters, where the electronic density is low, and thus the calculations have been performed under the BN approximation. To match both types of calculations, the GGA lifetime values for large vacancy clusters have been obtained by multiplying the corresponding BN value by the ratio of the bulk lifetimes calculated under the GGA and BN approximations ($\tau_{GGA}/\tau_{BN}$). The agreement between the calculated and 'matched' GGA values is excellent for mono- and divacancy.

The lifetime calculated from the relaxed configuration containing a monovacancy yields a value of 223 ps, which is only 1 ps shorter as compared with the unrelaxed structure. The slightly lower value accounts for the small atomic relaxation around a vacancy for the potential model used. The first nearest neighbours experience an inward relaxation of 0.3% whereas the second nearest neighbours relax outwards 0.2%. These results are qualitatively in agreement with the previous works of Hansen et al. [2] and Khanna et al. [4] for Mo; however, at least an atomic relaxation of -3% in the first neighbours is required to observe noticeable changes in the lifetime of a monovacancy in Mo. Our calculated values for Zn are in good agreement with the experimental lifetime value associated to monovacancies measured in electron irradiated Zn, $\tau$=220ps, [22] demonstrating that the atomic relaxation around a monovacancy predicted by the potential model is enough to explain the positron experiments.

The lifetime calculated for the relaxed divacancy is equal to 260 ps, which is only 1 ps shorter than the value obtained for the unrelaxed configuration. In this case the two first neighbours experience an inward relaxation of 1.1 and 1.4%; the second neighbours relax outwards 0.8%. As before, the relaxation is not large enough to give noticeable changes in the lifetime.

For the tri-vacancy there are two non-equivalent planar configurations (labelled *a* and *b* in fig. 1) that yield different results. The *a* configuration shows a large outward relaxation for the first neighbours whereas the second neighbours relax inward in the basal plane. As a result, the lifetime in the relaxed structure is 12 ps higher than the value in the unrelaxed defect. However, there is no difference between the unrelaxed and relaxed values for the tri-vacancy in configuration *b*. In this case the relaxation for the first neighbours out of the basal plane is also positive and about an order of magnitude lower than in the previously discussed *a* configuration. On the contrary, the second neighbours lying in the basal plane experience in both cases a negative relaxation of approximately the same magnitude (see table 1). These results confirm the importance of the first neighbours relaxation on the positron lifetime. As a further example not presented in the table, the lifetime for the relaxed 5-vacancy cluster is 2 ps shorter than the value calculated for the unrelaxed defect as a result of a 2% inward relaxation of the first neighbours. For the remaining configurations studied the first neighbours relaxation is below 0.7% and its effect on the lifetime is negligible.

The dependence of the lifetime values obtained from the relaxed configurations as a function of the vacancies contained in the cluster has been plotted in Fig. 4. It is observed, as expected, that the lifetime increases with the open volume associated to the defect and saturates to a fixed value close to 500 ps, as has been previously reported both theoretical and experimentally [23]. The effect of atomic relaxation is visible only for mono-, di- and type (a) trivacancy; for larger clusters lattice relaxation is not significant.

The results corresponding to the interstitial-type loops are collected in table 2. The calculated lifetime value for the O-loops is insensitive to the loop size and equals the bulk value. However, the loops formed by two interstitial layers exhibit values closer to the monovacancy lifetime, simply due to the presence of the double dislocation loop that increases the open volume associated to the defect in comparison with the single loop structure.

It is remarkable that the lifetime is strongly dependent of the loop structure and that in none of the cases it reaches the lifetime value for the monovacancy. It is thus inferred that the positron trapping at interstitial-type dislocation loops is strongly dependent of the loop atomic structure; the positron traps can range from shallow (O-loops) to vacancy-type traps (S-S and O-S loops); to improve their efficiency as positron traps, the O-loops have to absorb a vacancy. Particularly, for a 19-interstitials O-loop, the lifetime increases to 203 and 230 ps for vacancies in the sites labelled 1 and 2 respectively (see table 2 and fig. 2b).

Our results are in agreement with the theoretical calculations by Häkkinen et al. [3], who reported the existence of a variety of traps originated by different types of dislocations in fcc metals. They reported that the lifetime for a dislocation line in both Al and Cu is only one ps longer than in the bulk. Moreover, the presence of vacancies on the dislocation line increases the lifetime up to values characteristic of vacancies.

The same conclusions have been drawn by Kuramoto et al. for edge dislocations in Fe and Ni [7]. The O-loops in Zn follow the same behaviour. Thus, as a general feature, dislocation lines can be described as shallow traps for positrons whereas vacancies on the dislocation line or in its surroundings are responsible of the lifetime values currently assigned to dislocations. However, there is a point that deserves further comment. The lifetime obtained by Häkkinen et al. [3] for a vacancy-type dislocation loop formed on a {111}-plane in Al is 10% longer than the bulk lifetime but shorter than the lifetime for a vacancy. In the light of the previous results, one should expect a value close to the bulk as in the case of the interstitial-type O-loops. The reason of this value could be presumably attributed to a weak relaxation in Al that originates an open volume inside the loop sensed by the positron.

The comparison of the present data for positron lifetimes in defects in Zn with experimental positron measurements in electron irradiated Zn [22] allows us unambiguously to rule out the formation of three-dimensional vacancy clusters. Moreover, on the light of the present results, we can interpret the residual trapping signal observed after annealing to positron trapping at interstitial-type loops. The corresponding lifetime represents an average of trapping at the different kinds of loops and vacancies pinned in their surroundings.

## 4. Conclusions

The lifetime calculations carried out for several relaxed defect structures in Zn lead to the following conclusions:

- the effect of atomic relaxation in the positron lifetime associated to vacancy-type defects in Zn is significant only for mono-, di-; and type (a) trivacancies.

- the positron lifetime is very sensitive to the atomic structure of interstitial-type loops and it is increased by the presence of vacancies in the loop neighbourhood.

-the comparison of the theoretical calculations with experimental results allows to confirm the absence of three-dimensional vacancy clusters in electron irradiated Zn.

*Ackowledgment*. - This work has been supported by Ministerio de Educación y Cultura (Spain) under grant PB95-0374/95

Tables

Table 1 - Theoretical and experimental lifetime values in bulk Zn and in unrelaxed vacancy defects. The subscripts BN and GGA refer to the models as explained in the text. The calculations have been performed for an hexagonal lattice with $a$=0.266 nm and $c/a$=1.86. The positive (negative) relaxation values mean that atoms relax outwards (inwards) with respect to the defect centre.

|  | $\tau_{BN}$ (ps) | $\tau_{GGA}$ (ps) | $\tau_{exp}$ (ps) | $\Delta\tau$ (ps) | % relaxation 1st neighbours | % relaxation 2nd neighbours |
|---|---|---|---|---|---|---|
| Bulk | 139 | 158 | 152±1(*) | ------ | ------ | ------ |
| Monovacancy | 196 | 224 | 220(**) | 1 | -0.3 | 0.2 |
| Divacancy | 226 | 261 |  | 1 | -1.4/-1.1 | 0.8 |
| Trivacancy (a) | 258 | 305 |  | -12 | 9.8 | -1.0 |
| Trivacancy (b) | 258 | 305 |  | 0 | 0.6 | -0.9 |

(*) This value represents the mean of eight measurements performed in our laboratory in a well annealed 6N pure Zn sample with two different spectrometers having a resolution (FWHM) of 210 and 230 ps
(**) From ref. [21]

Table 2 - The calculated positron lifetimes for several interstitial-type loops. $V_1$ and $V_2$ denote respectively vacancies in first and second neighbour positions (see Section 2.1). $N_i$ refers to the number of single interstitials in the loop.

|  | $N_i$ |  | $\tau$ (ps) |
|---|---|---|---|
| O-loop | 7 | (fig 2(a)) | 158 |
| O-loop | 19 | (fig 2(a)) | 158 |
| O-loop + $V_1$ | 19 | (fig 2(b)) | 203 |
| O-loop + $V_2$ | 19 | (fig 2(b)) | 230 |
| S-S loop | 64 | (fig 3(a)) | 200 |
| O-T loop | 64 | (fig 3(b)) | 211 |

Figure captions

Fig.1  Schemes used for the construction of vacancy clusters in a basal plane projection. Full and open symbols denote lattice sites and vacancies respectively. Different types of symbols indicate different layers parallel to the basal plane.

Fig. 2  (a) Scheme used for the construction of interstitial-type O-loops; circles represent O-sites at z=c/4 that form 7- and 19-interstitial clusters. (b) The relaxed configuration of a 19-O type loop in a $\langle 1\bar{1}00\rangle$-type projection, showing six adjacent atom layers, denoted in the figure by different symbols. The faulted nature of the loop is clearly demonstrated. Sites marked 1 and 2 denote vacancies in first and second neighbour positions with respect to the loop (see text for explanation).

Fig. 3  The relaxed configuration of a S-S (a) and O-T type loop in a $\langle 1\bar{1}00\rangle$-type projection, showing four and six adjacent atom layers respectively, denoted in the figure by different symbols.

Fig. 4  The calculated positron lifetime as a function of the number of vacancies in the relaxed cluster; the bulk value has been marked as a reference.

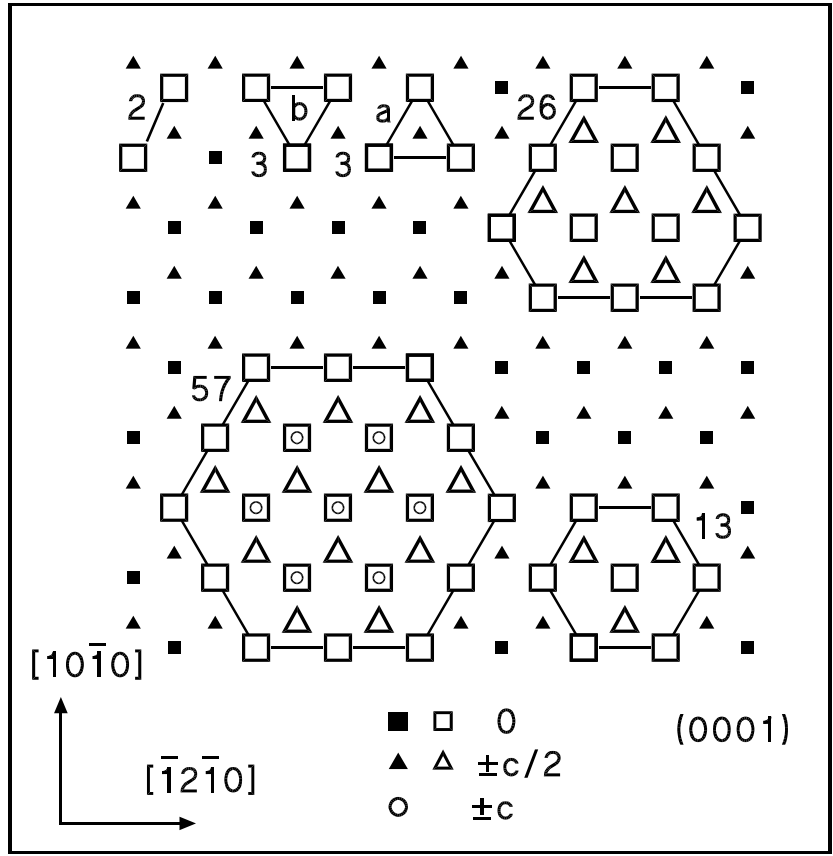

Fig. 1

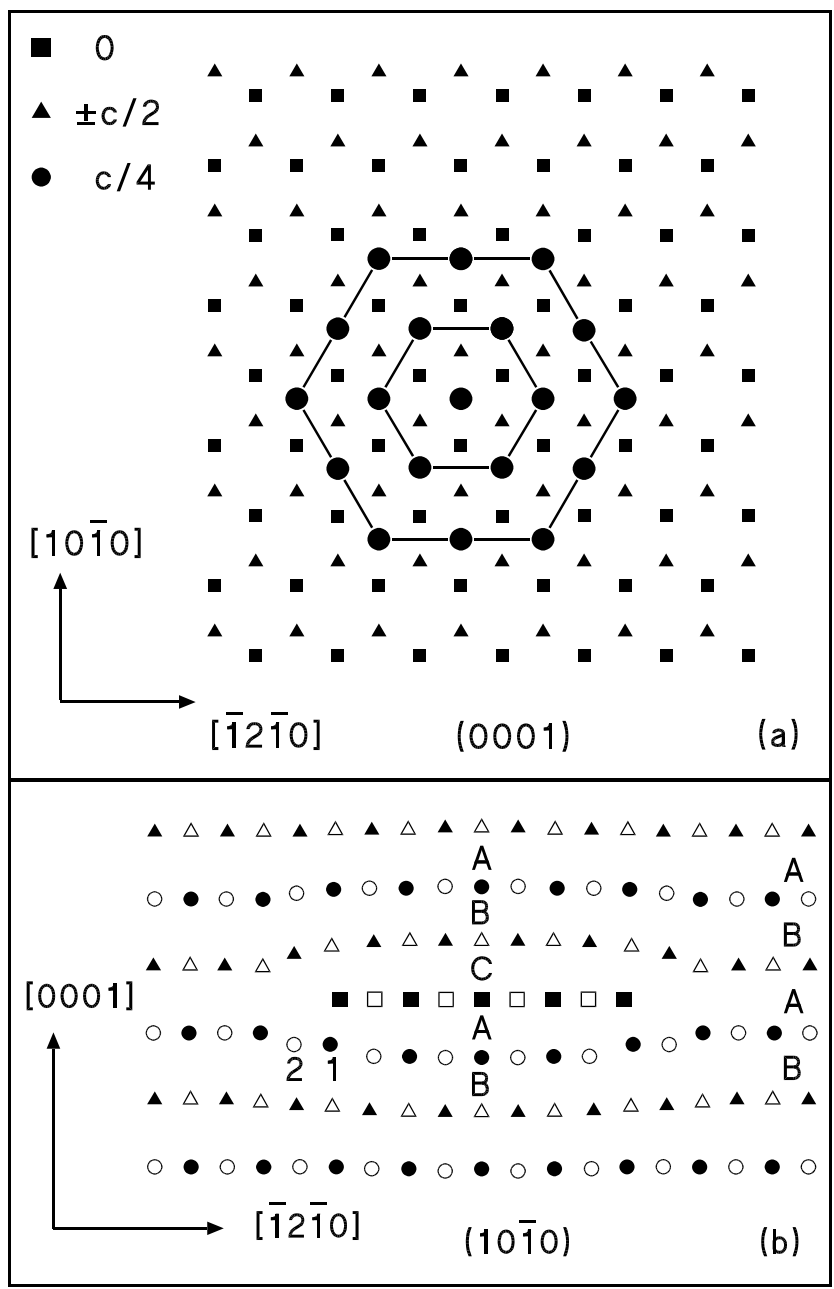

Fig. 2

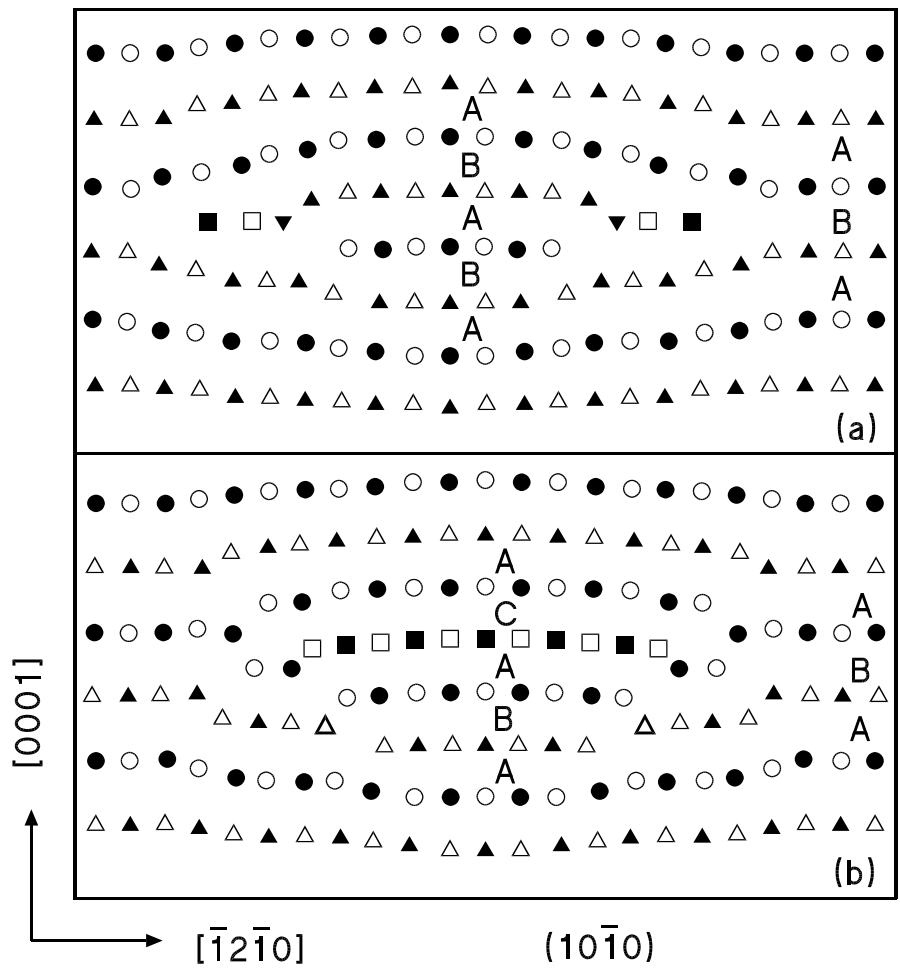

Fig. 3

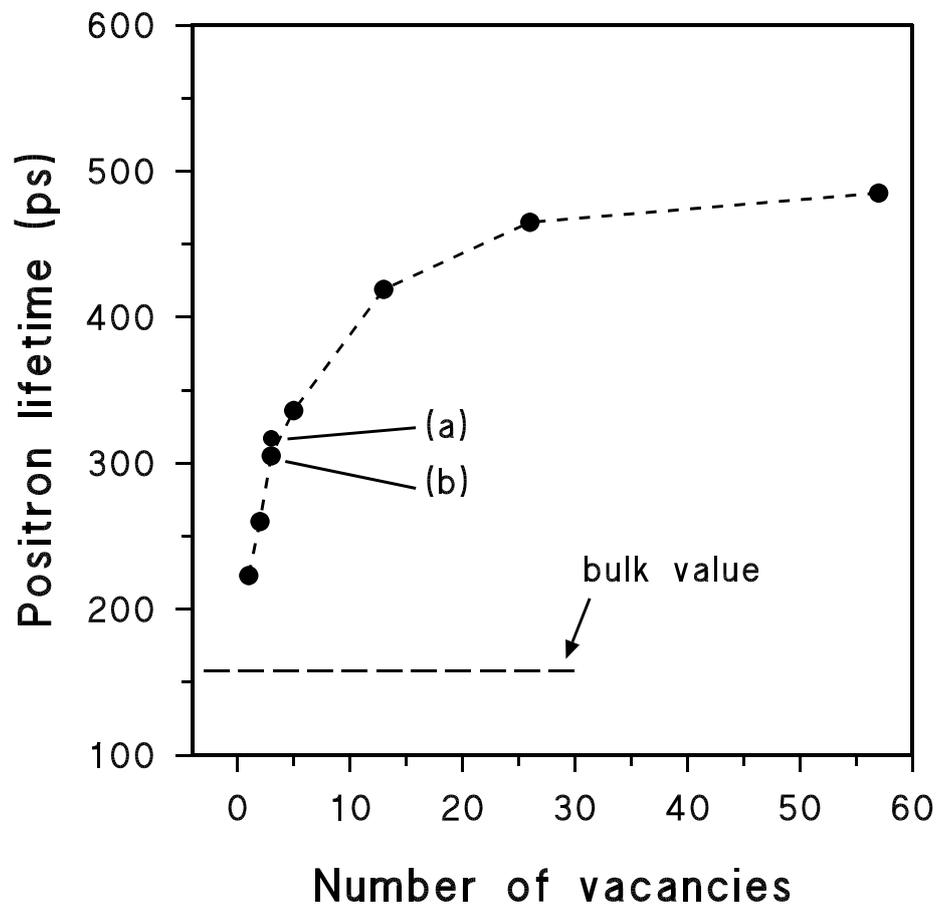

Fig. 4